\documentclass[twocolumn,showpacs,preprintnumbers]{revtex4}
\usepackage{graphicx}
\usepackage{amsmath}
\usepackage{amssymb}

\begin{document}

\title{New methods to determine the Hausdorff dimension of vortex loops in the $3DXY$ model}
\author{M. Camarda}
\author{F. Siringo}
\author{R. Pucci}
\affiliation{
Dipartimento di Fisica e Astronomia, Universit\`a di Catania,\\
and Lab. MATIS-INFM, and CNR-CNISM, Sez. Catania, and INFN, Sez. 
Catania,
64, Via S. Sofia, I-95123 Catania, Italy.}
\email{massimo.camarda@ct.infn.it}
\author{A. Sudb\o} 
\author{J. Hove}
\affiliation{Department of Physics, Norwegian University of Science and Technology, N-7491 Trondheim, Norway}

\begin{abstract}
  The geometric properties of critical fluctuations in the $3DXY$
  model are analyzed. The $3DXY$ model is a lattice model describing
  superfluids. 
   We present a \emph{direct} evaluation of the Hausdorff dimension
  $D_H$ of the vortex loops which are the critical fluctuations of the
  $3DXY$ model. We also present analytical arguments for why
  $\vartheta$ in the scaling relation $\eta_{\phi} + D_H = 2 +
  \vartheta$ between $D_H$ and the anomalous scaling dimension of the
  corresponding field theory, must be zero.
 \end{abstract}

\pacs{74.60.-w, 74.20.De, 74.25.Dw}
\maketitle
\section{Introduction}
It has been proposed that a second-order phase transition is
characterized by the breakdown of a generalized rigidity associated
with a proliferation of defect structures in the order parameter
\cite{Anderson_book}.  In the case of extreme type-II superconductors,
it has been explicitly demonstrated that the system undergoes a
continuous phase transition {\it driven by a proliferation of closed
  loops of quantized vorticity}
\cite{Tesanovic_blow_out,Nguyen,Nguyen_alpha_determination}.
In particular, it has been demonstrated that the connectivity of the
vortex-tangle (the tangle of topological defects of the system),
changes dramatically at the critical temperature of the system
\cite{Nguyen,Nguyen_alpha_determination}. In other words, it is
possible to describe the superconducting phase transition in terms of the proliferation of defect structures (viz. vortices and flux lines), which are singular phase fluctuations and determine the critical properties of the theory.  Using this argument, it is possible to relate the critical properties of the phase transition 
to the geometric properties of the loops \cite{Hove_Sudbo} at the
critical point.

The phenomenology of superconductivity usually starts with the
Ginzburg Landau (GL) model. The Hamiltonian for this model is
\begin{eqnarray}
\label{eq:G-L}
H(q,u_\psi)=m_\psi^2 \vert\psi\vert^2 +\frac{u_\psi}{2}
\vert\psi\vert^4 +\vert D_\mu\psi\vert^2+ \frac{1}{2} \left( \nabla
  \times \mathbf{A} \right)^2. 
\end{eqnarray}
Here $\psi = |\psi|e^{i\theta}$ is a complex matter field, 
coupled to a massless gauge field $\mathbf{A}$ through the 
minimal coupling $D_\mu = \partial_\mu - iqA_\mu$. $m_\psi$ 
is the mass parameter for the $\psi$ field, and $u_\psi$ is 
the self-coupling. This is a rich model which embodies many different aspects of superconductivity\cite{Mo:2002}, in this 
paper we will focus on the limit $\kappa \gg 1$ corresponding 
to extreme type-II superconductors. In this case the gauge 
field fluctuations can be ignored, and we are left with a 
neutral $|\psi|^4$ theory. For this theory it can be shown
that amplitude fluctuations in $\psi$ are innocuous\cite{Nguyen,Nguyen_alpha_determination}, and only the phase variables must be retained. When the resulting model is 
defined on a lattice, one obtains the 3DXY model.

The 3DXY model is a model for \emph{phase fluctuations}, and both the
smooth spin wave fluctuations, and the singular \emph{transverse}
fluctuations are accounted for. The transverse phase fluctuations are
defined by
\begin{eqnarray}
  \nabla \times \nabla \theta(\mathbf{r}) = 2 \pi \mathbf{n}(\mathbf{r})
  \label{vorticity}
\end{eqnarray}
where $\mathbf{n}(\mathbf{r})$ is the local {\it vorticity}. The
vortices are the critical fluctuations of the theory, which drive the
superfluid density to zero in the neutral case, and the magnetic
penetration length to infinity in the charged case. We have performed simulations directly on the phase degrees of freedom, 
and extracted the vortex content according to Eq. (\ref{vorticity}), alternatively it is possible to integrate 
out the spin wave degrees of freedom and retain only the 
vortex degrees of freedom \cite{Peskin}. The definition
Eq. (\ref{vorticity}) ensures that $\nabla \times \mathbf{n} = 0$
everywhere, hence the vorticity must be in the form of 
\emph{closed loops}.

Since the vortex loops are the critical fluctuations of the theory it
is natural to formulate a theory expressed in terms of these degrees
of freedom. In $d=3$ it is possible to start with the charged theory
Eq. (\ref{eq:G-L}) in a fixed-amplitude approximation and
derive\cite{Peskin} a field theory for the vortex loops. It
turns out that this \emph{dual} theory is a neutral $|\phi|^4$ theory.
Hence, the dual of a charged superfluid is a neutral superfluid and
vice versa, and furthermore the vortex loops of a neutral (charged)
superfluid are described by a field theory isomorphic to a charged
(neutral) superfluid. In this paper we will use the convention that
$\psi$ represents the original superfluid, and that $\phi$ is the
corresponding dual. Then $\phi$ will be the field theory for the
vortices of $\psi$.

The description in terms of loops is physically appealing. Firstly,
it highlights the physical meaning of the $|\phi|^4$ term which,
depending on the sign of $u_\phi$, represents a steric repulsion (as
in the case of type-II superconductors) or a steric attraction for
type-I superconductors with a first order transition\cite{Mo:2002}.
In the case of a neutral superfluid the vortices will interact
attractively through a long range interaction mediated by a dual gauge field, this will yield a vortex tangle more dense than a set of random loops\cite{Hove_Mo_Sudbo}.

A geometric interpretation of the critical point in terms of
proliferating geometric objects was given already in 1967
by M. E. Fisher \cite{Fisher:1967}. He considered droplets of 
one phase immersed in a background phase. Close to the critical point the distribution of droplet size was argued to behave as
\begin{equation}
  \label{Fisher}
  n(s) \propto \frac{1}{s^{\tau}} e^{-s \epsilon},\quad \epsilon \propto |T - T_c|^{\frac{1}{\sigma}},
\end{equation}
$n(s)$ is the mean number of droplets of ``mass'' $s$. The behavior
of Eq. (\ref{Fisher}) is governed by two critical exponents $\sigma$
and $\tau$, where the former governs the vanishing tension when
approaching the critical point and $\tau$ is related to the entropy of a droplet. These two exponents can be related to the six ordinary
critical exponents
\begin{xalignat}{3}
\nonumber
\alpha             &= 2 - \frac{\tau - 1}{\sigma}          &    \beta_{\mathsf{G}} &= \frac{\tau - 2}{\sigma}  &   \delta_{\mathsf{G}} &= 1 + \frac{\tau - 1}{d(\tau - 2)}  \\
\label{Gexp}
\eta_{\mathsf{G}}   &= 2 + \frac{d(\tau - 3)}{\tau - 1}     &     \nu              &= \frac{\tau - 1}{d\sigma} &  \gamma_{\mathsf{G}} &= \frac{3 - \tau}{\sigma}.
\end{xalignat}
The distribution $n(s)$ Eq. (\ref{Fisher}) is also used to describe
the cluster density close to the critical point in percolation, and the relations Eq. (\ref{Gexp}) can be easily derived from that
context as well\cite{Stauffer:1991:book}. The exponents in Eq. (\ref{Gexp})
have an index $\mathsf{G}$ to emphasize their \emph{geometric} origin. These exponents will in general not agree with those of the
underlying model. 
The Fortuin-Kasteleyn clusters\cite{Fortuin:1972} of the Q-state Potts
model is a special case where the exponents derived from $\tau$ and
$\sigma$ agree with those of the ordinary Potts model, i.e.
$\beta_{\mathsf{G}} = \beta, \gamma_{\mathsf{G}} = \gamma, \ldots$.
The vortices of the 3DXY model, which we study in this paper, constitute another special case, here the exponents derived from $\tau$ and $\sigma$ agree with the \emph{dual} of the initial theory. Since the dual of a neutral superfluid is a charged superfluid, the study of the vortices in the 3DXY model can actually be used to glean knowledge of the critical properties 
of a charged superfluid.  In particular, this can be used to 
relate the anomalous scaling dimension of the Ginzburg
Landau theory, $\eta_{\phi}$ to the fractal dimension of the 
vortex loops in the 3DXY model.

The paper is organized as follows. In section \ref{sec:HD_eta} 
we derive the relation between the fractal Hausdorff dimension $D_H$ of the loops and the anomalous dimension of the dual condensate, $\eta_\phi$. Sections \ref{sec:PercolatingSistems} 
and \ref{sec:3DXY} are devoted to presenting our
results from Monte Carlo simulations. As a benchmark, we 
determine the fractal dimension of percolation clusters in section
\ref{sec:PercolatingSistems}, and in section \ref{sec:3DXY} we
determine the fractal dimension of the vortex loops of the 3DXY
model. Finally, in Section \ref{sec:conclusions} we discuss our results.

\section{Relation between the Hausdorff and the anomalous dimensions}
\label{sec:HD_eta}
The anomalous dimension $\eta_\phi$ of the $\phi$ field is defined as
the critical exponent of the correlation function as follows
\begin{eqnarray}
\label{rel:correlation_function} 
G(\textbf{x},\textbf{y})=\langle\phi(\textbf{x})\phi^{\dag}(\textbf{y})\rangle.
\end{eqnarray}
This correlation function has the standard form at large 
distances at the critical point, namely
\begin{eqnarray}
G(x) \sim 1/x^{d-2+\eta_\phi}  \qquad  x \to \infty ,
\end{eqnarray} 
where $d$ the spatial dimension of the system.  
Particle-field duality dictates that this correlation 
function has a geometric interpretation, yielding the 
probability amplitude of finding any particle path 
connecting $\textbf{x}$ and $\textbf{y}$.  In the present 
work, the particle trajectories correspond to vortex loops
which can be deduced from the phase distribution of the matter field. It is essential that any vortex path connecting two points
$\textbf{x}$ and $\textbf{y}$ {\it must be part of a closed vortex
loop}, since only closed vortex loops provide the vortex paths in
this system.  To highlight the important properties of the probability
amplitude $P(\textbf{x},\textbf{y},N)$ of connecting two points
$\textbf{x}$ and $\textbf{y}$ by a continuous vortex path, we start by
focusing on the properties of the vortex loops.

It is known that for random loops in a lattice, the number of steps $N$,
defined as the number of occupied links, and the average distance $R_F
= \sqrt{\Delta_x^2+\Delta_y^2+\Delta_z^2}$, where $\Delta_i$ are the
coordinate variations of the trajectories, are related by
\begin{eqnarray}
\label{rel:Rf_N} 
R_F \sim  N^{1/D_H} \qquad  N \rightarrow \infty
\end{eqnarray} 
where $D_H$ is the Hausdorff dimension \cite{def:fractal_dimension}.\\
Since these vortex are the critical fluctuations of the theory their
average size $R_F$ is related to the correlation length $\xi$ of the
field $\phi$ so that we can set $\xi \sim R_F$ 
\begin{eqnarray}
  \label{rel:xi_N}
  \xi \sim N^{1/D_H}.
\end{eqnarray}
With this we can now relate the correlation function $G(x,y)$ 
with the Hausdorff dimension $D_H$ by taking advantage of the 
fact that in every second order transition the system is scale invariant at $T=T_c$. We can, then, write the following scaling Ansatz for the probability
amplitude $P(\textbf{x},\textbf{y},N)$
 \begin{eqnarray}
\label{rel:Scaling-Ansaltz} 
P(x,y,N) &\propto& P(\vert r \vert, N) \propto \frac{1}{\xi^{d}} F\left(\frac{\vert r \vert}{\xi}\right) \nonumber \\ &\propto& N^{-d/D_H} F\left(\frac{\vert r \vert}{N^{1/D_H}}\right)
\end{eqnarray}
where $P(x,y,N)$ is the probability of finding a loop of length $N$
connecting the points $x$ and $y$. The probability of coming back
to the starting point is generally given by $P(x,x,N)$. For some
models like self avoiding walks (SAW) and random walks in $d \ge 3$
this probability is vanishing, and the scaling function $F(z)$ has the limiting behavior
\begin{equation}
  \label{Scaling:Limit}
  \lim z \to F(z) = z^{\vartheta}, \qquad  \vartheta > 0.
\end{equation}
In the case of vortex \emph{loops}, which are closed, we clearly must have $\vartheta = 0$. This is discussed in more detail in Appendix \ref{App:closed}.  At the critical point the two-point correlation function $G(r)$, Eq.(\ref{rel:correlation_function}), scales with an anomalous dimension $\eta_\phi$. At the same time 
we know that, by the definition of $P(x,y,N)$ together with Eq.(\ref{rel:Scaling-Ansaltz}):
\begin{eqnarray}
\label{rel:G-P} 
G(x,y)=\sum_N P(x,y,N) = \sum_N N^{-d/D_H} F\left(\frac{\vert r \vert}{N^{1/D_H}}\right).
\end{eqnarray}
If we focus on the long loop regime ($N \gg 1$), we can replace the
summation with an integral and obtain
\begin{eqnarray}
\label{rel:Integrazione-G} 
G(x,y) \sim \int dn \: n^{-d/D_H} F\left(\frac{\vert x-y\vert}{n^{1/D_H}}\right)=  \frac{C}{\vert x-y \vert^{d-D_H}}.
\end{eqnarray}
Therefore, we can relate the anomalous dimension $\eta_\phi$ to the
fractal dimension of the vortex loops, $D_H$, as follows
\begin{eqnarray}
\label{rel:Anom_Dim-Frac_Dim} 
\eta_\phi + D_H = 2.
\end{eqnarray}
We can also relate $D_H$ and, indirectly $\eta_\phi$, to another
critical exponent, $\tau$, which is related to the density of loops
in the system. From the definition of loop density we have
\begin{eqnarray}
\label{rel:Density-Probability} 
D(N) \propto \frac{1}{N}\sum_x P(x,x,N)
\end{eqnarray}
here, $P(x,x,N)$ is the probability of finding a closed loop of length $N$ and $D(N)$ is the mean number of loops of length $N$.  Near $T_c$ the loop distribution will then take the form
\begin{eqnarray}
\label{rel:Density-Scaling} 
D(N) \sim N^{-\tau}.  
\end{eqnarray} 
If the system is homogeneous, all the contributions in
Eq.(\ref{rel:Density-Probability}) are equal.  Thus, picking 
an arbitrary $x$ we find
\begin{eqnarray}
\label{rel:Probability-tau} 
P(x,x,N) \propto \frac{N}{V} D(N) \propto N^{1-\tau}.
\end{eqnarray}
Finally, combining Eq.(\ref{rel:Scaling-Ansaltz}), Eq.(\ref{rel:Probability-tau}) and Eq.(\ref{rel:Rf_N})
we arrive at the following relations
\begin{eqnarray}
\label{rel:Tau-D_H} 
\frac{d}{D_H} &=& \tau-1 \\
\label{rel:Anom_Dim-tau} 
\eta_\phi &=& 2 - D_H= 2 - \frac{d}{\tau-1}.
\end{eqnarray}
It can be seen from Eq.(\ref{rel:Anom_Dim-Frac_Dim}) and
Eq.(\ref{rel:Anom_Dim-tau}) that the anomalous dimension of the 
dual condensate can be determined either through the 
determination of $\tau$ or $D_H$, the former being 
statistically easier to evaluate. However, $\eta_\phi$ is 
quite sensitive to $\tau$ 
[$\partial \eta_\phi/ \partial \tau = d/(\tau - 1)^2 \sim 1.5 $].

The purpose of the present work is to carry out a {\it direct
evaluation} of $D_H$ which facilitates a check of the 
scaling law Eq. (\ref{rel:Tau-D_H}). For this reason, the 
first part of this study has been devoted to test their 
validity and the validity of the algorithms used in the case 
of a percolation transition, since it has been extensively 
studied in the literature and can be easily implemented.
\begin{figure}[t]
\centering
\includegraphics[height=0.9\columnwidth,angle=-90]{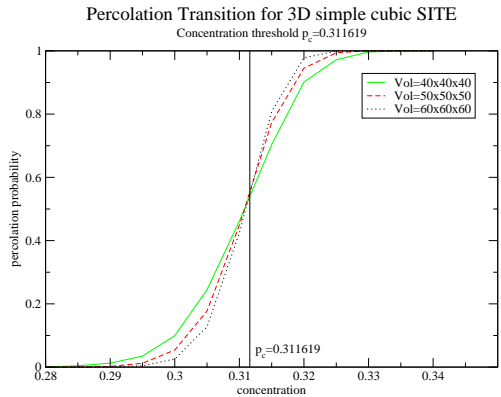}
\caption{(Color online) Percolation probability of sites as a function of concentration $p$ for a simple 3D cubic system. The concentration at which percolation takes place is given by the common 
inflection point of the different curves for increasing 
lattice size.}
\label{fig:p_c_site.ps}
\end{figure}

\begin{figure}[t]
\centering
\includegraphics[height=0.9\columnwidth,angle=-90]{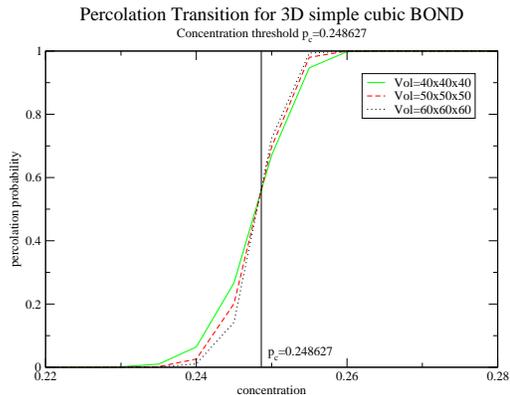}
\caption{(Color online) Same as Fig.\ref{fig:p_c_site.ps}, but for a bond percolation}
\label{fig:p_c_bond.ps}
\end{figure}
Before we start, let us return to the points made above concerning the HT-graph expansion of the partition function and two-point correlation functions\cite{Prokof,Janke}. Suppose one were to compare the Hausdorff dimension of physical vortex loops \cite{Hove_Sudbo} with the Hausdorff dimension of the sort of 
open ended, self-avoiding walks that one would consider in a HT-graph expansion of the two-point correlation function \cite{Prokof}. We may compute the loop-distribution function 
for vortex loops, $D_{\rm{loop}}(N) \sim N^{-\tau_{\rm{loop}}}$, with the distribution of ``closed'' graphs in
self-avoiding walks $D_{\rm{SAW}} (N) \sim N^{-\tau_{\rm{SAW}}}$.  
A self-avoiding walk of length $N$ is defined as ``closed'' if 
one after $N$ steps ends up on a lattice point which is nearest neighbor to the starting point \cite{Prokof,Janke}, see also 
Fig. 1 of \cite{Janke}. From Fig. 1 in Ref. \cite{Janke}, it 
is immediately clear that the stop criterion for defining a ``closed'' self-avoiding walk, if applied to real physics 
vortex loops, would mean that one would discard a tail in 
the distribution function involving large vortex loops.  This 
is so since one could connect the remaining last bridge by an arbitrarily long and complicated vortex path, not just over 
the shortest link. (In fact, these two point {\it were} 
connected by an arbitrarily long and complicated path, and 
there is no reason to close the path only by the shortest distance). This would lead to an overestimate for $\tau_{\rm{loop}}$, since removing large loops means
that $D(N)$ drops too much as a function of $N$. An 
overestimate of $\tau$ would lead to an underestimate of 
$D_H$, cf.  Eq. (\ref{rel:Tau-D_H}). In the scaling 
relation $\eta_{\phi} + D_H = 2$, it is crucial to compute 
$D_H$ from the real physical vortex loops of the system. If 
one underestimates $D_H$ by for instance considering the 
fractal Hausdorff dimension of other objects than real 
physical vortex loops, this might lead one to erroneously 
conclude that one needs to introduce an additional exponent $\vartheta >0$ and modify the scaling relation to 
$\eta_{\phi} + D_H = 2 -\vartheta$. While $\vartheta > 0$ 
is needed in the case of self-avoiding walks, this is
not so for vortex loops due to the topological constraint 
${\bf  \nabla} \cdot {\bf n} =0$.

\section{Results for the percolating systems}
\label{sec:PercolatingSistems}
Percolation theory is used to describe a variety of natural physical
processes where disorder is an essential ingredient. Applications
range from spontaneous magnetization of diluted ferromagnets,
formation of polymer gels, electrical conductivity of amorphous
semiconductors and many others\cite{Stauffer:1991:book}.

In link (site) percolation the links (sites) of a lattice are
occupied independently with probability $p$, and then the resulting
clusters of connected links (sites) are analyzed. For small $p$ we
will only have small clusters, and the there will be no path
connecting the edges of the system. For large $p$ there will be large clusters comparable to the entire system, and the system 
can sustain a current from edge to edge. The transition from an insulator to metal is a second order (geometric) phase 
transition at a critical value $p_c$.

Figs.\ref{fig:p_c_site.ps} and \ref{fig:p_c_bond.ps} show our results for the percolation probability, i.e. the probability 
that there is at least one cluster spanning the whole system, 
for sites and bond percolation respectively. We have found 
$p_c = 0.311619$ and $p_c = 0.248627$, respectively. These 
values agree reasonably well with existing results \cite{pc_perc_determination}. Clusters have been identified 
using the Hoshen-Kopelman (HK) \cite{HK} labeling algorithm and 
up to $10^4$ configurations were considered in the measurements.

We next focus on the calculation of the fractal dimension of
percolation clusters at the critical point, and compare with results
from the literature \cite{D_H_percol_determination}. 
Formally the fractal dimension can be defined from the \emph{box
counting technique} \cite{def:fractal_dimension}. The fractal 
object is covered with boxes of linear size $l$, and we count 
the number of boxes needed for complete coverage. The fractal dimension is inferred from the variation in total number of 
boxes $N(l)$ with box size $l$ as follows
\begin{eqnarray}
\label{def:D_H} 
N(l) \propto {\lim_{l \to 0}}\;\; l^{-D_H}.
\end{eqnarray}
Defining $l$ on a lattice is difficult since the lattice spacing $a$
is a fixed dimensionless constant.  Instead, we can relate $l$ to
$1/R_F$, which is equivalent to considering loops of different size
$R_F$ as the same loop, but at a different length scale.  This relation
is \emph{unique} provided that the different loops can be embedded in
a box $\Delta_x \Delta_y \Delta_z$ of the same shape, i.e.
\begin{eqnarray}
\label{rel:proporzioni_costanti} 
\frac{\Delta_i}{\Delta_j}=C_{ij} \qquad \forall \quad i \neq j \qquad i,j \; \in \;\; [1,3].
\end{eqnarray}
where $C_{ij}$ are constants enforcing a \emph{constant proportion
  constraint} ($\tt{CPC}$). This implies that loops have to be divided
into groups, each one characterized by the actual values of the two
independent ratios $C_{ij}$, say $C_{xy}$ and $C_{xz}$, and
Eq.(\ref{def:D_H}) applies \emph{separately} to each group. In the
current paper we have only considered cubic boxes with $\Delta_x =
\Delta_y = \Delta_z$.

From Eq.(\ref{rel:Rf_N}), it can be seen that there is a relation
between the dimension of the cluster (i.e.  the number of cells) and
its radius (i.e. the dimension of the box containing it).  Thus, we
first relate the number $N$ of occupied cells by the loops to their
length $R_F$ using an embedding box of fixed shape $\Delta_x =
\Delta_y = \Delta_z$. Although this approach is very close in spirit to
the definition Eq. (\ref{def:D_H}) the method fails miserably giving
$D_H=1.329(8)$, which deviates strongly from the literature results
$D_H=2.5230$\cite{D_H_percol_determination}.

The problem, we believe, is that this method (later called $\tt{Method1}$) is affected by two
strong restrictions. Firstly, every box must scale in all the three
directions in the same way (because of the $\tt{CPC}$), i.e. all the
boxes have to be cubes. Secondly, the extraction of the box can be
mistaken if the cluster extends beyond the borders of the lattice, so that boxes touching the borders are excluded.  For these two reasons the statistics of $N(R_F)$ in $\tt{Method1}$ is extremely poor, and a correct value of $D_H$ cannot be found.

Another way to extract $D_H$ is to consider the largest cluster for
every configuration, and assume that the box containing this cluster
consists of the entire lattice. When the linear size of the lattice is varied this gives the fractal dimension from
\begin{equation}
  N^{\ast} \propto L^{D_H},  
\end{equation}
where $N^{\ast}$ is the size of the largest cluster and $L$ is the
linear extent of the lattice. Using this approach we have found $D_H = 2.52$, which is consistent with the value $\tau = 2.189$.

Hence, the study of the percolation transition has been useful to
first test the simulation and the algorithms used for the extraction
of the geometrical properties and, ultimately, it has provided a test of Eq.(\ref{rel:Tau-D_H}).  We can now proceed to the more difficult task of determining the Hausdorff dimension for the ensemble of vortex loops generated by thermal transverse phase fluctuations of the $3DXY$ model.
\section{Results for the 3DXY model}
\label{sec:3DXY}
The Hamiltonian for the case of a neutral condensate in the London approximation (i.e. neglecting amplitude fluctuations) is given
by
\begin{eqnarray}
\label{eq:3DXY}
H(q=0,\theta)=- J \sum_{\langle i,j \rangle} \cos(\theta_i-\theta_j).
\end{eqnarray}
Eqs.(\ref{rel:Tau-D_H}) and (\ref{rel:Anom_Dim-Frac_Dim}) are valid
only at the transition point where the system is scale invariant.
Therefore the simulations must be restricted to the case $T=T_c$. 
\begin{figure}[t]
\centering
\includegraphics[height=\columnwidth,angle=-90]{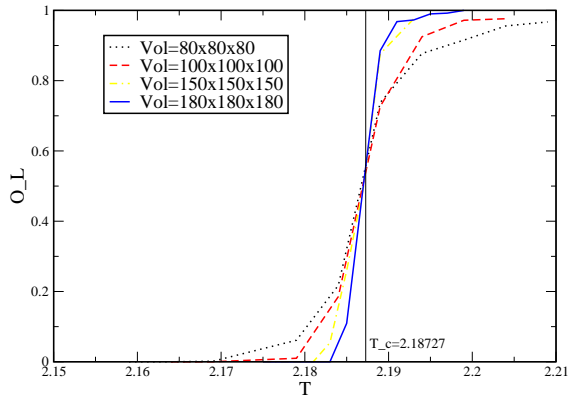}
\caption{(Color online) Percolation probability $O_L$ as function of temperature $T$ for the $3DXY$ model at different 
lattice sizes.  The critical temperature is given by the common inflection point of the different curves 
for increasing lattice size. The value obtained is $T_c=2.18727$}
\label{fig:Tc_3DXY}
\end{figure}
We have used the standard Metropolis algorithms for lattice sizes of
$L=80,100,150,180$, all the runs had a hot start and 5000 sweeps were discarded before measurements were made. Vortices are found
identifying singularities in the phase configuration according to Eq. (\ref{vorticity}). The temperature found ($T=2.18727$, cf.
Fig.\ref{fig:Tc_3DXY}) is close to, but different from the
thermodynamical one, $T_c=2.20184$ \cite{T_c_3dxy}.  Variations 
with the dimension of the system are still present, and it cannot 
be excluded that in the thermodynamical limit the two values merge
together. More extensive simulations along with a finite-size scaling analysis will be necessary to provide this question with 
a more satisfactory answer.

For the time being, since any $T \geqslant T_c$ is better than 
$T < T_c$ to test Eq.(\ref{rel:Tau-D_H}), all the following simulations will be done at the fixed temperature of $T_{c}=2.20184$. Now that the temperature has been set, $\tau$ 
and $D_H$ can be determined separately and their values compared 
to the ones available in the literature.

\begin{figure}[t]
\centering
\includegraphics[height=\columnwidth,angle=-90]{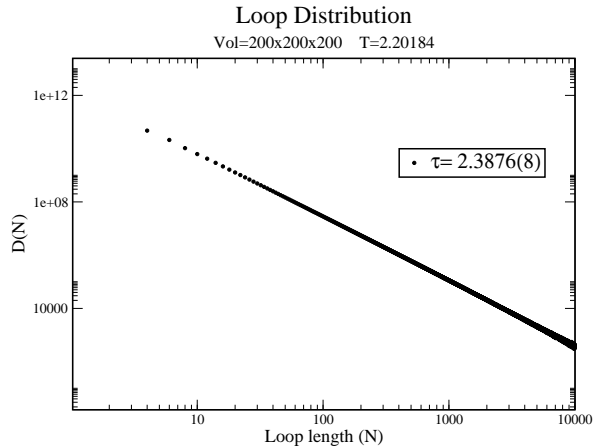}
\caption{Vortex-loop distribution $D(N) \sim N^{-\tau}$ as 
a function of perimeter length $N$ at the temperature 
$T=2.18727$, as explained in the text. }
\label{fig:D_N_3dxy.ps}
\end{figure}

All the simulations were performed on a system of fixed size $L=200$
using up to 40 2Ghz Pentium4.  The Monte Carlo runs had a cold start
and $10^4$ sweeps were discarded for thermalization before taking
measurements over $4 \cdot 10^5$ configurations. The loops in the
system are identified with a modified version of the Hoshen-Kopelman
algorithm which resolves vortex crossings randomly. The exponent
$\tau$ for the loop distribution has been determined earlier as
$\tau=2.310(3)$ \cite{Nguyen_alpha_determination, Hove_Sudbo}. The
results obtained for $\tau$ are shown in Fig.\ref{fig:D_N_3dxy.ps}.

We now turn to the fractal dimension $D_H$. The first interesting fact is that the method used to extract the pair (length, box) used in the
percolation transition does not work any more, giving unphysical
values for $D_H$. On the other hand, the more correct method of
extracting the exact box containing the loops yields a value in
agreement with Eq.(\ref{rel:Tau-D_H}).

There is also a topological reason for which we can expect to have
different statistics for the two systems (and therefore we need to
apply two different methods). When extracting the clusters in a
percolating system, \emph{all} connected sites (or bonds) are
considered part of the same object, i.e. belonging to the same
cluster. In general, this will lead to fewer but larger clusters
in the system, this is particularly the case at the transition point, where the cluster tension vanishes. In the case of vortex-loops, on the other hand, two intersecting loops are connected  with $50\%$ probability, which directly influences 
the value of $\tau$ which is smaller in the case of percolation.

\subsection{\tt{Method1}}
\begin{figure}[t]
\centering
\includegraphics[height=\columnwidth,angle=-90]{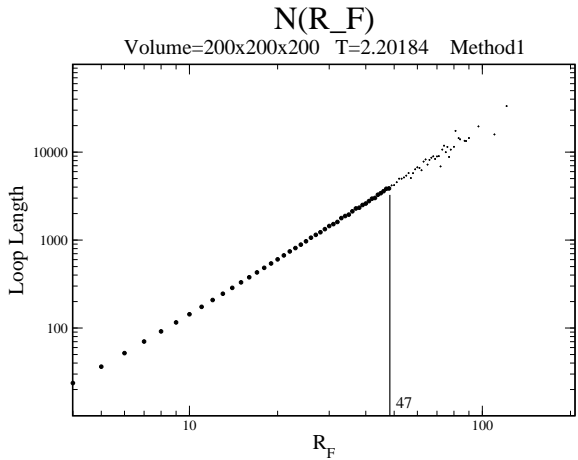}
\caption{Loop length $N(R_F)$ as a function of the characteristic size $R_F$. The best fit has been performed only on the point between $4\leq R_F\leq 47$.}
\label{fig:D_H_3dxy_Method1}
\end{figure}

We have tested $\tt{Method1}$, and the extracted value of
$D_H=2.168(6)$ (Fig.\ref{fig:D_H_3dxy_Method1}) is in good agreement
with the expected one of $D_H \simeq 2.162(1)$, as calculated using
Eq.(\ref{rel:Tau-D_H}), and the value of $\tau=2.3876(8)$ previously
obtained (Fig.\ref{fig:D_N_3dxy.ps}).

\begin{figure}[t]
\centering
\includegraphics[height=\columnwidth,angle=-90]{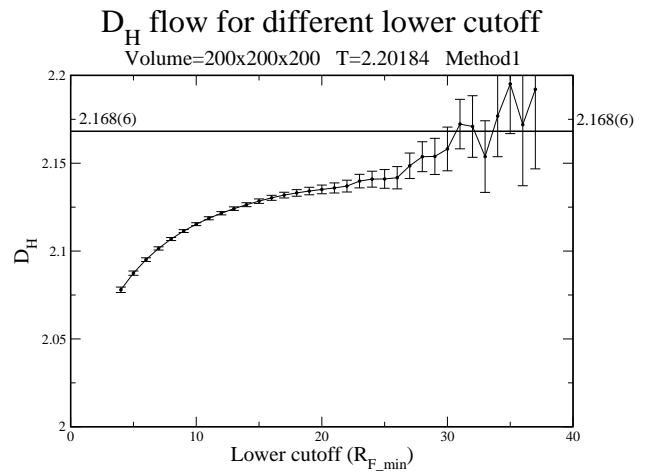}
\caption{Variation of $D_H$ with the lower cutoff. The saturation expected in the continuum limit is present but strongly effected 
by fluctuations. The finale value of $D_H$=2.168(6) is obtained using a weighted fit over the last eight points}
\label{fig:Saturation_Method1}
\end{figure}

Eq.(\ref{rel:Rf_N}) is assumed to hold only in the limit $N \to
\infty$, so that small deviations from linearity are expected for
small $N$ where the discrete form of the lattice becomes important.
Therefore, in order to reduce these lattice effects, we have varied
the lower cutoff on $R_F$ systematically. Results are shown in
Fig.\ref{fig:Saturation_Method1}, which is characterized by a
`saturation' in the long-loop regime. To obtain a reliable value of
$D_H$ this saturation region has to be reached, but, although this is the case, there are still considerable fluctuations.


For this reason and owing to the results obtained for percolating
systems, we have considered two new methods in order to determine
values of the Hausdorff dimension $D_H$ with better precision.  The
main limitation in \texttt{Method1} is the limited statistics, in an
attempt to improve on this situation we have devised two additional
methods to calculate $D_H$. Both methods rely on `relaxation' of the
$\tt{CPC}$ constraint. 

The strongest statistical limitation on $\tt{Method1}$ is that the
loops have to be divided into groups and that only comparison within
each group are allowed. The consequence of this restriction can be
directly seen in Fig.\ref{fig:D_H_3dxy_Method1}, where the longest
loop considered has a linear dimension of $\sim 47$, to be compared
with the linear dimension of the lattice of $200$.  Clearly then, the longest loops present in the lattice (the ones for which
Eq.(\ref{rel:Rf_N}) holds) are excluded from the final statistics
obtained.

\subsection{{\tt Method2}}
\begin{figure}[t]
\centering
\includegraphics[height=\columnwidth,angle=-90]{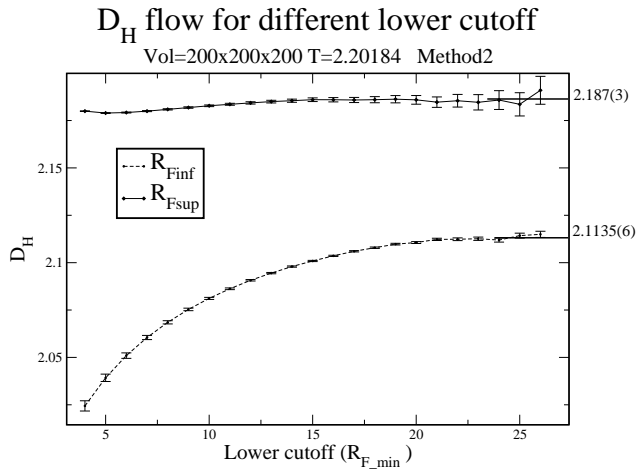}
\caption{Dependence of $D_{H-}$ and $D_{H+}$ on the lower cutoff ($R_F$).  The deviation between the two values is still too large even in the long-loop regime.  The finale value of $D_H$=2.14(7) is obtained averaging over the final values for $D_{H}^-$ and $D_{H}^+$}
\label{fig:Saturation_Method3}
\end{figure}

The first of these two new methods ($\tt{Method2}$ below) determines a lower and an upper bound for $R_F$.  To see how 
this can be done let us consider a box containing a loop with 
the condition
$\Delta_x=\Delta_y=\Delta$ and $\Delta_z=\Delta+C$ where $C$ is some
constant, and $\Delta_x,\Delta_y,\Delta_z$ the dimensions of the box
containing the loop in the three directions.  We immediately observe
that, since $C$ is constant, there is a violation of the $\tt{CPC}$
\begin{eqnarray}
\label{} 
\frac{\Delta_z}{\Delta_x}=\frac{\Delta+C}{\Delta} \neq \frac{\Delta^\prime+C}{\Delta^\prime}=\frac{\Delta^\prime_z}{\Delta^\prime_x}.
\end{eqnarray}
However, the error is of the order of $C/\Delta$, so in the limit of
large boxes ($\Delta \gg C$) this approximation is exact. We can then define the following loop size $R_F$ as follows
\begin{eqnarray}
\label{def:Volumi} 
R_F     &=& \sqrt{\Delta_x^2+\Delta_y^2+\Delta_z^2} \\
R_{F}^- &=& \Delta\sqrt{3} \\
R_{F}^+ &=& (\Delta+C)\sqrt{3},
\end{eqnarray}
where $R_F^{-}$ and $R_F^{+}$ are lower and upper bounds for $R_F$.
By definition, we have $R_{F}^- < R_F < R_{F}^+$ for every $\Delta$,
so that $D_{H}^- < D_H < D_{H}^+$ $\forall$ $\Delta$.  Moreover,
\begin{eqnarray}
\lim_{\Delta \rightarrow \infty} D_{H}^- =\lim_{\Delta \rightarrow \infty} D_{H}^+ = \lim_{\Delta \rightarrow \infty} D_{H}^+ = D_H 
\label{def:limiti_equiv} 
\end{eqnarray}
Then, $D_{H}^-$ and $D_{H}^+$ represent a lower and upper bond,
respectively, for the Hausdorff dimension, thus allowing a direct
estimate of the error in $D_H$.  As can be seen from
Fig.\ref{fig:Saturation_Method3}, the separation between the two
values obtained, $D_{H-}$ and $D_{H+}$, is still of the order of the
fluctuations in Fig. \ref{fig:Saturation_Method1}, hence there is no
significant improvement in the $D_H$ determination.

\subsection{{\tt Method3}}
We finally turn to the last method ($\tt{Method3}$ below) which will prove to be the best one to extract a precise value for $D_H$.
\begin{figure}[t]
\centering
\includegraphics[height=0.9\columnwidth,angle=-90]{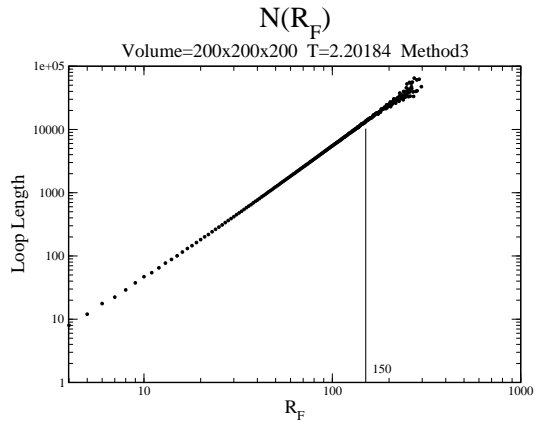}
\caption{Loop length $N(R_F)$ as a function of the characteristic size $R_F$.  The best fit has been performed 
on the points between $4\leq R_F\leq 150$, to be compared with that of $\tt{Method1}$, in Fig.\ref{fig:D_H_3dxy_Method1}.}
\label{fig:D_H_3dxy_Method4}
\end{figure}
In contrast to what is the case for {\tt{Method1}} and {\tt{Method2}},
in {\tt{Method3}} there is no constraint on the proportions of the box
containing the loops, since it is based on the approximation that,
considering a large number of configuration, the different values for
$D_H$ will eventually compensate. As can be seen from
Fig.\ref{fig:Saturation_Method4}, this method gives much better
statistics both in the small- and large-loop regimes. Moreover, the
longer loops considered in the statistic, having a linear dimension of
$\sim 150$ (cf. Fig.\ref{fig:D_H_3dxy_Method4}), although greater than
the one considered with $\tt{Method1}$ and $\tt{Method2}$, is still
smaller then the linear dimension of the lattice. Hence, finite size
effects should not substantially influence the final value of $D_H$
obtained.
\begin{figure}[t]
\centering
\includegraphics[height=\columnwidth,angle=-90]{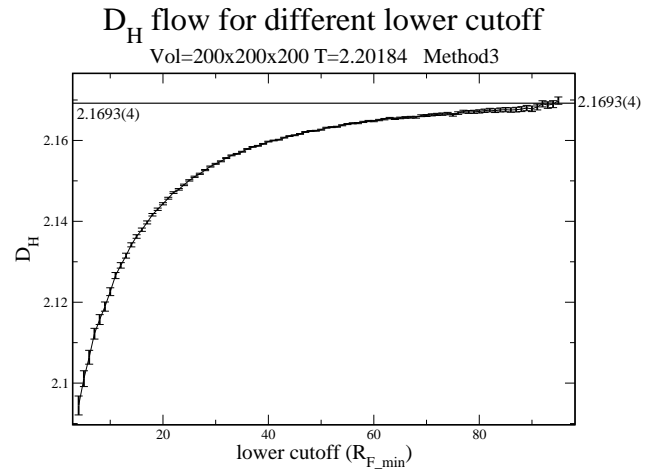}
\caption{Dependence of $D_H$ on the lower cutoff ($R_{F\_min}$).  The saturation region is, as can be seen, much clearer compared with previous methods.  The finale value of $D_H=2.1693(4)$ is obtained using a weighted fit over the last three points}
\label{fig:Saturation_Method4}
\end{figure}

\section{Conclusions}
\label{sec:conclusions}
The results obtained for the different methods are summarized in
Tab.\ref{Tab:risultati_finali}.

All methods are in agreement with Eq.(\ref{rel:Tau-D_H}) and
consistent with each other, although they have different accuracies.
As predicted by ref.\cite{string_statistic}, finite size effects are
present both in the short and long-loops regime
(Fig.\ref{fig:D_N_3dxy.ps}), but there is a clear monotonic behavior
towards the continuum limit
(Figs.\ref{fig:Saturation_Method1},\ref{fig:Saturation_Method3} and
\ref{fig:Saturation_Method4}). The best results are obtained 
from {\tt  Method 3}, which yields a final value for the 
anomalous dimension of the condensate $\eta_\phi=-0.1693(4)$. 
This is consistent with both perturbative renormalization 
group calculations \cite{Folk,Herbut,Kang,Halperin} and 
previous simulations
\cite{Hove_Sudbo,Nguyen_alpha_determination,Hove_Mo_Sudbo}.

In ordinary local quantum field theories without gauge fields, one can
prove that $\eta_\phi$ must be greater or equal to zero.  In contrast,
in gauge theories the proof of a non-negative $\eta_\phi$ is not
applicable due to the non gauge-invariant form of the correlation
function \cite{Nogueira,Kleinert_1,Kleinert_2}.  Moreover, a negative
value of $\eta_\phi$ implies a fractal dimension larger than that of
Brownian random walks, which means that the current lines are
self-seeking.  This point has been thoroughly discussed in
\cite{Hove_Mo_Sudbo}.  Indeed a $D_H> 2$ (corresponding to $\eta_\phi
< 0$) is necessary for the possibility of the existence of a phase
transition driven by a vortex-loop unbinding even in the presence of a
finite external magnetic field \cite{Nguyen,Tesanovic_blow_out}.  This
value is different from the one obtained for the 3DXY model, showing
that the gauge fluctuations of the Abelian Ginzburg-Landau model (dual
$3DXY$) modify the critical behavior, so that the Abelian gauge model
and the $\phi^4$ model \emph{belong to two different universality
  classes}.

\begin{table}
\caption{Values of $\tau$ and $D_H$, it can be seen that Eq.(\ref{rel:Tau-D_H}) is satisfied for all methods, and that all the values of $D_H$ extracted are consistent with each other. 
  The last two columns are a check of internal consistency.}
\label{Tab:risultati_finali}
\centering
\begin{tabular}{|c|c|c||c|c|}
\hline 
 & $D_H$       & $\tau$ & $\eta_{\phi} = 2 - \frac{d}{\tau - 1}$ & $\eta_{\phi} = 2 - D_H$\\ 
\hline 
$\tt{Method1}$ & $2.168(6)$ & 2.3876(8)  & -0.162(1)   & -0.168(6) \\ 
\hline                                               
$\tt{Method2}$ & $2.14(7)$ & 2.3876(8)  & -0.162(1)   & -0.14(7) \\ 
\hline                                               
$\tt{Method3}$ & $2.1693(4)$ & 2.3876(8) & -0.162(1)  & -0.1693(4) \\ 
\hline 
\end{tabular} 
\end{table}

\begin{acknowledgments}
  One of us (MC) acknowledges partial financial support from MIUR,
  through Progetto InterLink II00085947, from CSFNSM, Catania, Italy,
  and the HPC-EUROPA project (RII3-CT-2003-506079), with the support
  of the European Community - Research Infrastructure Action under the
  FP6 Structuring the European Research Area Programme. The work of AS
  was funded by the Research Council of Norway, Grant Nos. 157798/432,
  158518/431, and 158547/431 (NANOMAT), and 167489/V30 (STORFORSK).
  The figures were made by MC.  The large-scale Monte-Carlo
  simulations presented in the main body of the paper, were   performed
  by MC. AS acknowledges useful discussions and communications 
  with Z. Tesanovi\'c, N. V. Prokof'ev, and B. V. Svistunov. 
\end{acknowledgments}

\appendix
\section{Vortex loops and  Eq. (\ref{Scaling:Limit})}
\label{App:closed}
The quantity
\begin{eqnarray}
\label{eq:Prokof}
\lim_{r \to 0} F\left(\frac{\vert r \vert}{\bar{\xi}}\right) = \lim_{z\to 0} F(z) = z^\vartheta \qquad  \vartheta >0
\end{eqnarray}
is a scaling function for the probability
of traversing the distance between two points along
a continuous vortex path, and holds for an ensemble of 
\emph{both closed and open self-avoiding walks} \cite{Gennes_polimeri}. Eq.(\ref{eq:Prokof}) is often 
used in polymer physics to describe the statistical 
properties of polymer tangles.  The statement that
$\vartheta > 0$ is certainly not true for the statistical 
properties of vortex loops, which are, 
\emph{by construction}, closed and not self avoiding, so 
we have $P(x,x,\bar{N}) \ne 0$ and $\vartheta = 0$
\cite{ReplyHove}. The main physical difference between 
vortex loops in a superconductor and a tangle of polymers, 
is that while polymers can start and end inside a system, 
this is not so for vortex loops. They must either close 
on themselves or alternatively thread the entire
system.  If we denote by ${\bf n}$ the local vorticity arising from
transverse phase fluctuations, defined by Eq. (\ref{vorticity}), then we have the constraint ${\bf \nabla} \cdot {\bf n} = 0$ everywhere inside the superfluid (and superconductor).  We reemphasize that this in particular means that {\it each and 
every vortex path in the system must be part of a closed 
vortex loop}. No paths of vortices can be open-ended.  Hence, 
for the former we can have $\vartheta > 0$ while for the 
latter $\vartheta =0$. To emphasize this further, let us
consider the physical meaning of a positive $\vartheta$ \cite{Janke}. In the problem of self-avoiding walks, 
$\vartheta > 0$ is an exponent that governs the asymptotic 
number of open-ended paths of length $N$ at the critical 
point, via the relation $z_N \sim N^{\vartheta/D_H}$.
We also emphasize that in this paper and in previous works, 
we have \cite{Nguyen,Nguyen_alpha_determination,Hove_Sudbo} exclusively been dealing with the geometrical properties of vortex-loop paths at the critical point in superfluids
and superconductors. \\

Let us also make another remark which is relevant in this context. It is known \cite{Book_Kleinert}, that the partition function of 
the $3DXY$ model may be expanded in so-called high-temperature (HT)
graphs, which are isomorphic to the vortex-loop gas of a
superconductor. In the case of charged superconductor the fluctuating
gauge field screens the Coulomb interactions, and the resulting loop
interactions are short range\cite{Hove_Sudbo}. Hence, for the
purposes of studying the critical {\it thermodynamics} of a
superconductor, one might as well use such a HT expansion of the
partition function, instead of summing over all configurations of
closed vortex loops of the system. In other words, at the level of the {\it partition function}, one can elevate the HT graphs to objects
that are equivalent to real physical topological defects in the
superconducting order parameter, i.e. closed vortex loops. A similar
property is known for the two-dimensional Ising model, where one at
the level of the partition function can proceed with a HT expansion
and identify a formal equivalence between the closed HT graphs and the topological defects of the theory, namely closed lines in a $2D$ plane connecting domains of oppositely directed Ising spins.

However, when we compute correlation functions, as we need to do for
computing the probabilities we have discussed above, the connection
between physical closed vortex loop paths and the graphs of the HT
expansion {\it for the correlation functions}, is far less obvious.
When one utilizes the HT expansion to compute the correlation function
$\langle\phi(\textbf{x})\phi^{\dag}(\textbf{y})\rangle$, one has to
consider {\it open-ended graphs starting at ${\bf x}$ and ending at
  ${\bf y}$} \cite{Janke}. One may, if one wishes, define a closed
such path as the path at which the endpoint of the walk for the first
time reaches a lattice point which is the nearest neighbor of the
starting point, see for instance Fig. 1 of \cite{Janke}. One may
further go on to discuss the geometrical properties of such open-ended
(and ``closed'') HT graphs involved in computing the two-point
correlation function in a HT expansion. However, the graphs that one
then ends up with studying, have nothing to with the closed vortex
paths that are the topological defects of the superconductor, and
which were the objects under consideration in our previous work
\cite{Hove_Sudbo}.  (The fractal structure of HT graphs in $O(N)$
models in two spatial dimensions, as well as the fractal structure of
spin clusters and domain walls in the two-dimensional Ising model, has
recently been investigated in detail
\cite{Janke_PRL2005,Janke_PRB2005}.)  The Hausdorff dimension of
such{\it correlation function HT graphs} will not be the same as the
Hausdorff dimension of the closed vortex loops of a superconductor. Such open-ended graphs violate the constraint that the vortex-loop 
paths {\it must} respect, namely that they must form continuous closed paths and cannot start or end inside the system.

\bibliography{massimo}

\begin{thebibliography}{10}

\bibitem{Anderson_book}
P.~W. Anderson, {\em Basic Notions in Condensed Matter} (Addison-Wesley,
  Redwood City, California, 1984).

\bibitem{Tesanovic_blow_out}
Z.~Tesanovi\'c, Phys. Rev. B {\bf 59},  6449  (1999).

\bibitem{Nguyen}
A.~K. Nguyen and A.~Sudb\o{}, Euro. Phys. Lett. {\bf 46},  780  (1999).

\bibitem{Nguyen_alpha_determination}
A.~K. Nguyen and A.~Sudb\o{}, Phys. Rev. B {\bf 60},  15307  (1999).

\bibitem{Hove_Sudbo}
J.~Hove and A.~Sudb\o{}, Phys. Rev. Lett. {\bf 84},  3426  (2000).

\bibitem{Mo:2002}
S.~Mo, J.~Hove, and A.~Sudb{\o}, Phys. Rev. B {\bf 65},    (2002).

\bibitem{Peskin}
M.~Peskin, Ann. Phys. (N.Y.)  (1978).

\bibitem{Hove_Mo_Sudbo}
S.~Mo, J.~Hove, and A.~Sudb\o{}, Phys. Rev. Lett. {\bf 85},  2368  (2000).

\bibitem{Fisher:1967}
M.~E. Fisher, Rep. Prog. Phys. {\bf 30},  615  (1967).

\bibitem{Stauffer:1991:book}
D.~Stauffer and A.~Aharony, {\em Introduction to Percolation Theory} (Taylor \&
  Francis, London, new Fetter Lane 11, 1991).

\bibitem{Fortuin:1972}
C.~M. Fortuin and P~.W. Kasteleyn, Physica {\bf 57},  536  (1972).

\bibitem{def:fractal_dimension}
B.~B. Mandelbrot, {\em The Fractal Geometry of Nature} (Freeman, San
  Francisco,California, 1982).

\bibitem{Prokof}
N.~V. Prokof'ev and B.~V. Svistunov, Physics Review Letters {\bf 96},  219701
  (2006).

\bibitem{Janke}
W.~Janke and A.~M.~J. Schakel, cond-mat/0508734 (unpublished).

\bibitem{pc_perc_determination}
C.~D. Lorenz and R.~M. Ziff, Phys. Rev. E {\bf 57},  230  (1998).

\bibitem{HK}
J.~Hoshen and R.~Kopelman, Phys. Rev. B {\bf 14},  3438  (1976).

\bibitem{D_H_percol_determination}
M.~B. Isichenko, Rev. Mod. Phys. {\bf 64},  961  (1992).

\bibitem{T_c_3dxy}
Y.-H. Li and S.~Teitel, Phys. Rev. B {\bf 40},  9122  (1989).

\bibitem{string_statistic}
D.~Austin, E.~J. Copeland, and R.~J. Rivers, Phys. Rev. D {\bf 49},  4089
  (1994).

\bibitem{Folk}
R.~Folk and Yu. Holovatch, J. Phys. A {\bf 29},  3409  (1996).

\bibitem{Herbut}
I.~F. Herbut and Z.~Tesanovi\'c, Phys. Rev. Lett. {\bf 76},  4588  (1996).

\bibitem{Kang}
J.~S. Kang, Phys. Rev. D {\bf 10},  3455  (1974).

\bibitem{Halperin}
B.~I. Halperin, T.~C. Lubensky, and S.~K. Ma, Phys. Rev. Lett. {\bf 32},  292
  (1974).

\bibitem{Nogueira}
F.~S. Nogueira, Phys. Rev. B {\bf 62},  14559  (2000).

\bibitem{Kleinert_1}
H.~Kleinert and F.S.Nogueira, Nucl. Phys B {\bf 651},  361  (2003).

\bibitem{Kleinert_2}
H.~Kleinert and A.~M.~J. Schakel, Phys. Rev. Lett. {\bf 90},  97001  (2003).

\bibitem{Gennes_polimeri}
P.~G. de~Gennes, {\em Scaling Concepts in Polymer physics} (Cornell University
  Press, London, 1979), Chap.~I.

\bibitem{ReplyHove}
J.~Hove and A.~Sudbo, Physics Review Letters {\bf 96},  219702  (2006).

\bibitem{Book_Kleinert}
H.~Kleinert, {\em Gauge Fields in Condensed Matter} (World Scientific,
  Singapore, 1989), Vol.~1. Superflow and Vortex Lines.

\bibitem{Janke_PRL2005}
W.~Janke and A.~M.~J. Schakel, Phys. Rev. Lett. {\bf 95},  135702  (2005).

\bibitem{Janke_PRB2005}
W.~Janke and A.~M.~J. Schakel, Phys. Rev. E {\bf 71},  036703  (2005).

\end{thebibliography}
\bibliographystyle{mprsty}
\end{document}